\documentclass[aps, twocolumn, showpacs]{revtex4}
\usepackage{amssymb}
\usepackage{amsmath}
\usepackage{epsfig}
\usepackage{bm}

\begin{document}

%




\title{Revisiting the hopes for scalable quantum computation}                           

\author{M.I. Dyakonov}

\affiliation{Laboratoire Charles Coulomb, Universit\'e Montpellier II, CNRS, France}



\begin{abstract}

The hopes for scalable quantum computing rely on the ``threshold theorem'': once the error 
per qubit per gate is below a certain value, the methods of quantum error correction allow 
indefinitely long quantum computations. The proof is based on a number of assumptions, which 
are supposed to be satisfied \textit{exactly}, like axioms, e.g. \textit{zero} undesired interactions between qubits, etc. However in the physical world no continuous quantity can be \textit{exactly} zero, 
it can only be more or less small.  Thus the ``error per qubit per gate'' threshold must be 
complemented by the required precision with which each assumption should be fulfilled. 
This issue was never addressed. In the absence of this crucial information, the prospects of scalable quantum computing remain uncertain.

\end{abstract}

\maketitle

The idea of quantum computing is to store information in the values of $2^N$ 
complex amplitudes describing the wavefunction of $N$ two-level systems (qubits), 
and to process this information by applying unitary transformations (quantum gates), 
that change these amplitudes in a precise and controlled manner \cite{steane1}.
The value of $N$ needed to have a useful machine is estimated as $10^3$ or more. Note that  even $2^{1000}\sim 10^{300}$ is much, much greater than the number of protons in the Universe.

Since the qubits are always subject to various types of noise, and the gates cannot be perfect, it is widely recognized that large scale, i.e. useful, quantum computation is impossible without implementing  error correction. This means that the $10^{300}$ continuously changing quantum amplitudes of the grand wavefunction describing the state of the computer must closely follow the desired evolution imposed by the quantum algorithm. The random drift of these amplitudes caused by noise, gate inaccuracies, unwanted interactions, etc., should be efficiently suppressed.

Taking into account that all possible manipulations with qubits are not exact, it is not obvious at all that error correction can be done, even in principle, in an analog machine whose state is described by at least $10^{300}$ continuous variables. Nevertheless, it is generally believed (for example, see \cite{arda}) that the prescriptions for fault-tolerant quantum computation \cite{shor1,presk,got, steane2} using the technique of error-correction by encoding \cite{shor2,steane3} and concatenation (recursive encoding) give a solution to this problem. By active intervention, errors caused by noise and gate inaccuracies can be detected and corrected during the computation. The so-called ``threshold theorem'' \cite{ben-or,kitaev,knill} says that, once the error per qubit per gate is below a certain value estimated as $10^{-6} - 10^{-4}$, indefinitely long quantum computation becomes feasible. 

Thus, the theorists claim that the problem of quantum error correction is resolved, at least in principle, so that physicists and engineers have only to do more hard work in finding the good candidates for qubits and approaching the accuracy required by the threshold theorem \cite{alif, knill1}.

However, as it was clearly stated in the original work (but largely ignored later, especially in presentations to the general public, Ref. \cite{knill1} is one example) the mathematical proof of the threshold theorem is founded on a number of assumptions (axioms):\\

1. Qubits can be prepared in the $|00000...00\rangle$ state. New qubits can be prepared on demand in the state $|0\rangle$,

2. The noise in qubits, gates, and measurements is uncorrelated in space and time,

3. No undesired action of gates on other qubits,

4. No systematic errors in gates, measurements, and qubit preparation,

5. No undesired interaction between qubits,

6. No ``leakage'' errors,

7. Massive parallelism: gates and measurements are applied simultaneously to many qubits,

and some others.\\

While the threshold theorem is a truly remarkable mathematical achievement, one would expect that the underlying assumptions, considered as axioms, would undergo a close scrutiny to verify that they can be reasonably approached in the physical world. Moreover, the term ``reasonably approached'' should have been clarified by indicating with what precision each assumption should be fulfilled. So far, this has never been done (assumption 2 being an exception \cite{correlation, staudt}), if we do not count the rather naive responses provided in the early days of quantum error correction \cite{presk2,comment1, presk4}.

It is quite normal for a theory to disregard small effects whose role can be considered as negligible. But not when one specifically deals with errors and error correction. A method for correcting \textit{some} errors on the assumption that other (unavoidable) errors are \textit{non-existent} is not acceptable, because it uses fictitious ideal elements as a kind of golden standard \cite{dyakonov1}.

Below are some trivial observations regarding manipulation and measurement of continuous variables. Suppose that we want to know the direction of a classical vector, like the compass needle. 

First, we never know exactly what our coordinate system is. We choose the $x,y,z$ axes related to some physical objects with the $z$ axis directed, say, towards the Polar Star, however neither this direction, nor the angles between our axes can be defined with an infinite precision. Second, the orientation of the compass needle with respect to the chosen coordinate system cannot be determined exactly. 

So, when we say that our needle makes an angle $\theta = 45^o $ with the $z$ axis, we understand that $\cos \theta$ is not exactly equal to the irrational number $1/\sqrt{2}$, rather it is somewhere around this value within some interval determined by our ability to measure angles and other uncertainties. We also understand that we cannot manipulate our needles perfectly, that no two needles can ever point exactly in the same direction, and that consecutive measurements of the direction of the same needle will give somewhat different results. 

In the physical world, continuous quantities can be neither measured nor manipulated exactly. In the spirit of the purely mathematical language of the quantum computing literature, this can be formulated in the form of the following\\

\textbf{Axiom 1.} No continuous quantity can have an exact value.

\textit{Corollary.} No continuous quantity can be exactly equal to zero.\\

To a mathematician, this might sound absurd. Nevertheless, this is the unquestionable reality of the physical world we live in \cite{comment3}. Note, that \textit{discrete} quantities, like the number of students in a classroom or the number of transistors in the on-state, can be known exactly, and \textit{this} makes the great difference between the digital computer and the hypothetical quantum computer \cite{onoff}.

Axiom 1 is crucial whenever one deals with continuous variables (quantum amplitudes included). Each step in our technical instructions should contain an indication of the needed precision. Only then the engineer will be in a position to decide whether this is possible or not. 

All of this is quite obvious. 

Apparently, things are not so obvious in the magic world of quantum mechanics. There is a
widespread belief that the $|1\rangle$ and $|0\rangle$ states ``in the computational basis'' are 
something absolute, akin to the on/off states of an electrical switch, or of a transistor in a digital circuit, but with the advantage that one can use quantum superpositions of these states. It is sufficient to ask: ``With respect to {\it which} axis do we have a spin-up state?'' to see that there is a serious problem with such a point of view. 

It should be stressed once more that the coordinate system, and hence the computational basis,
cannot be exactly defined, and this has nothing to do with quantum mechanics. Suppose that, again, we have chosen the $z$ axis towards the Polar Star, and we measure the $z$-projection of the spin with a Stern-Gerlach beam-splitter. There will be inevitably some (unknown) error in the alignment of the magnetic field in our apparatus with the chosen direction. Thus, when we measure some quantum state and get $(0)$, we never know exactly to what state the wavefunction has collapsed. Presumably, it will collapse to the spin-down state with respect to the (not known exactly) direction of the magnetic field in our beam-splitter. However, with respect to the chosen $z$ axis (whose direction is not known exactly either) the wavefunction will always have the form $a|0\rangle+b|1\rangle$, where, hopefully, the unknown $b$ is small, $|b|^2\ll 1$. Another measurement with a similar instrument, or a consecutive measurement with the same instrument will give a different value of $b$. 

Quite obviously, the unwanted admixture of the $|1\rangle$ state is an error that \textit{cannot be corrected}, since (contrary to the assumption 1 above) we can never have the standard \textit{exact} $|0\rangle$ and $|1\rangle$ states to make the comparison. 

Thus, with respect to the consequences of imperfections, the situation is quite similar to what we have in classical physics. The classical statement ``the exact direction of a vector is unknown'' is translated into quantum language as ``there is an unknown admixture of unwanted states''. The pure state $|0\rangle$ can never be achieved, just as a classical vector can be never be made to point \textit{exactly} in the $z$ direction, and for the same reasons, since quantum measurements and manipulations are done with classical instruments.

Clearly, the same applies to {\it any} desired state. Thus, when we contemplate the ``cat state'' $(|0000000\rangle + |1111111\rangle)/\sqrt{2}$, we should not take the $\sqrt{2}$ too seriously, and we should understand that \textit{some} (maybe very small) admixture of e.g. $|0011001\rangle$ state must be necessarily present.\\

{\it Exact quantum states do not exist. Some admixtures of all possible states to any desired state are unavoidable.}\\

This fundamental fact described by Axiom 1 (nothing can be \textit{exactly} zero!) should be taken into account in any prescriptions for quantum error correction.

At first glance, it may seem that there \textit{are} possibilities for achieving a desired state with an arbitrary precision. Indeed, using nails and glue, or a strong magnetic field, we can fix the compass needle so that it will not be subject to noise. We still cannot determine exactly the orientation of the needle with respect to our chosen coordinates, but we can take the needle's direction as the $z$ axis.  However: 1)we cannot align another fixed needle in exactly the same direction and 2)we cannot use fixed needles in an analog machine, to do this, they must be detached to allow for their free rotation.

Quite similarly, in the quantum case we can apply a strong enough magnetic field to our spin at a low enough temperature, and wait long enough for the relaxation processes to establish thermodynamic equilibrium. Apparently, we will then achieve a spin-down  $|0\rangle$ state with any desired accuracy (provided there is \textit{no interaction} with other spins in our system, which is hardly possible). 

However ``spin-down'' refers to the (unknown exactly) direction of the magnetic field at the spin location. Because of the inevitable inhomogeneity of the magnetic field, we cannot use the direction of the field at the spin location to define the computational basis, since other spins within the same apparatus will be oriented slightly differently. Moreover, if we want to manipulate this spin, we must either switch off the magnetic field (during this process the spin state will necessarily change in an uncontrolled manner), or apply a resonant ac field at the spin precession frequency, making the two spin levels degenerate in the rotating frame. The high precision acquired in equilibrium will be immediately lost. 

Likewise, an atom at room temperature may be with high accuracy considered to be in its ground state. Atoms at different locations will be always subject to some fields and interactions, which mix the textbook ground and excited states. Also, such an atom is not yet a two-level system. In order for it to become a qubit, we must apply a resonant optical field, which will couple the ground state with an excited state. The accuracy of the obtained states will depend on the precision of the amplitudes, frequencies, and duration of optical pulses. This precision might be quite sufficient for many applications, but certainly it can never be \textit{infinite}.

Abstractions are intrinsic to Mathematics, and using them is probably the only way to develop a theoretical understanding of the physical world. However, when we specifically deal and try to fight with imperfections, noise, and errors, we should be extremely vigilant about mixing the abstractions and the physical reality, and especially about \textit{attributing} our abstractions, like exact quantum states, $\sqrt 2$, decoherence free subspaces, etc. to the physical reality \cite{comment4}.

Of course, \textit{if} the assumptions underlying the threshold theorem are approached with a high enough precision, the prescriptions for error-correction could indeed work. So, the real question is: \textit{what} is the required precision with which each assumption should be fulfilled to make scalable quantum computing possible?

How small should be the undesired, but unavoidable: interaction between qubits, influence of gates on other qubits \cite{comment5}, systematic errors of gates and measurements \cite{comment1}, leakage errors, random and systematic errors in preparation of the initial $|0\rangle$ states? Quite surprisingly, not only is there no answer to these most crucial questions in the existing literature, but they have never even been seriously discussed! Obviously, this gap must be filled, and the ``error per qubit per gate'' threshold must be complemented by indicating the required precision for each assumption.

Until this is done, one can  only speculate about the final outcome of such a research. The optimistic prognosis would be that some additional threshold values $\epsilon_1, \epsilon_2 ...$ for corresponding precisions will be established, and that these values will be shown neither to depend on the size of the computation nor to be extremely small. In this case, the dream of factorizing large numbers by Shor's algorithm might become reality in some distant future.

The pessimistic view is that the required precision must increase with the size of computation, most probably in an exponential manner, and this would undermine the very idea of quantum computing. 

Classical physics gives us some enlightening examples regarding attempts to impose a prescribed evolution on quite simple continuous systems.  For example, consider some number of hard balls in a box. At $t=0$ all the balls are on the left side and have some initial velocities. We let the system run for some time, and at $t=t_0$ we simultaneously reverse all the velocities. Classical mechanics tells us that at $t=2t_0$ the balls will return to their initial positions in the left side of the box. Will this ever happen in reality, or even in computer simulations?

The known answer is: Yes, \textit{provided} the precision of the velocity inversion is exponential in the number of collisions during the time $2t_0$. If there is some slight noise during the whole process, it should be exponentially small too. Thus, if there are only 10 collisions, our task is difficult but it still might be accomplished. But if one needs 1000 collisions, it becomes impossible, not because Newton laws are wrong, but rather because the final state is strongly unstable against very small variations of the initial conditions and very small perturbations. 

This classical example is not directly relevant to the quantum case (see Ref. \cite{gutz} for the relation between classical and quantum chaos).  However it might give a hint to explain why, although some beautiful and hard experiments with small numbers of qubits have been done (see Ref. \cite{experiments} for recent results with 3 to 8 qubits), the goal of implementing a concatenated quantum error-correcting code with 50 qubits (set by the ARDA Experts Panel \cite{arda} for the year 2012) is still nowhere in sight. 

There are two recurrent themes in discussions of the perspectives for scalable quantum computing. One of them is: ``Because there are no known fundamental obstacles to such scalability, it has been suggested
that failure to achieve it would reveal new physics'' \cite{knill1}. An alternative suggestion is that such a failure would reveal insufficient understanding of the role of uncertainties, and the inconsistency of a theory of error correction that carelessly replaces some \textit{small} quantities by \textit{zeros} \cite{comment6}.

The other one consists in directly linking the possibility of scalable quantum computing to the laws of Quantum Mechanics, so that we are forced to either admit or reject both things together:  ``The accuracy threshold theorem for quantum computation establishes that scalability is achievable provided that the currently accepted principles of quantum physics hold and that the noise afflicting a quantum computer is neither too strong nor too strongly correlated'' \cite{presk1, comm}.

Obviously, one can have full confidence in the principles of Quantum Mechanics, which are confirmed by millions of experimental facts, and at the same time have doubts about a theory of fault-tolerance which considers some unavoidable errors as non-existent.

In summary, the proof of the threshold theorem is founded on a number of assumptions that are supposed to be fulfilled exactly. Since this is not possible, an examination of the required precision with which these assumptions should hold is indispensable. The prospects of scalable quantum computing crucially depend on the results of such a study.


\begin{thebibliography}{}
\bibitem{correlation} Many publications were devoted to the study of different noise models in the context of quantum error correction, see Ref. \cite{staudt} for a review, and it was shown that assumption 2 can be somewhat relaxed by allowing for certain types of noise correlations.
\bibitem{comment1} There is not and never will be a single device dealing with continuous quantities that makes \textit{zero} systematic errors. Moreover, for reasons that are not yet well understood, all electronic devices, even the most precise that we have, the atomic clock, suffer from the so-called flicker or 1/f noise. The parameters of the device slowly but chaotically change in time, and the longer we wait the more changes we see.
\bibitem{onoff} In accordance with Axiom 1, there \textit{is} some current in the off-state. However because of the enormous value of the on/off ratio this is not a problem.
\bibitem{comment4} As far as we know, the postulates of Quantum Mechanics are true. However, they are true in the same sense as is true the statement ``The diagonal of a unit square is equal to $\sqrt{2}$''. It would be very naive to think that this literally applies to some physically \textit{real} unit square which we can deal with.  Similarly, the \textit{exact} $|0\rangle$ state is a mathematical abstraction that has no place in reality. Just as the $\sqrt{2}$ diagonal, it can be only \textit{approached with a certain limited precision}. Another mathematical abstraction is ``arbitrary accurately'' \cite{knill1}. This notion does not exist in the vocabulary of a physicist or an engineer.
\bibitem{comment5} Due to the required massive parallelism, thousands of gates, which in practice are electromagnetic pulses, will be applied simultaneously, so that the quantum computer will resemble a huge microwave oven. It must be a rather difficult problem for the future quantum engineer to exclude the unwanted action of gates on other qubits.
\bibitem{comment6} A similar ``aproximation'' will allow the hard balls in a box to always return to their initial positions after velocity inversion.
\bibitem{comm} It should have been added: ``and also that the assumptions on which the theorem relies are satisfied exactly''.
\end{thebibliography}
\end{document}